\documentstyle[prb,aps,twocolumn,epsfig]{revtex}
\draft
\begin{document}
\title{Band structure analysis of the conduction-band mass anisotropy in
6H and 4H SiC}
\author{Walter R. L. Lambrecht and Benjamin Segall}
\address{Department of Physics, Case Western Reserve University\\
Cleveland, OH-44106-7079}
\date{\today}
\maketitle
\begin{abstract}
The band structures of 6H SiC and  4H SiC
calculated by means of the full-potential
linear-muffin-tin orbital method are used to determine
the effective mass tensors for their conduction-band minima.
The results  are shown to be consistent with recent
optically detected cyclotron resonance measurements which find the ratio of
cyclotron masses for ${\bf B}\perp {\bf c}$ to ${\bf B}\parallel {\bf c}$
to be larger (smaller) than unity for the 6H (4H) polytype. However,
contrary to previous suggestions,  appreciable anisotropies in the
c-plane are found. For 6H SiC, a strong dependencies on band-filling
is predicted because of the occurence of a double well minimum along the
ML-axis. The calculated mass tensors for 3C and 2H  are also reported.
\end{abstract}
\narrowtext

The \hbox{technological potential} of SiC for high-tempera-ture,
high-power and high-frequency
electronic devices as well as the intrinsic scientific interest
 of the polytypism of SiC
have been recognized for several decades.
Only recently, however, has it become possible to obtain electronic
grade quality single crystal single polytype material.
Recently, it was found that 4H SiC exhibits
a much smaller (and opposite) anisotropy of the electron mobility
with respect to the c-axis  than does 6H SiC.\cite{Schaffer,Schadt}
It has been  unclear up to now whether this could be attributed
to a corresponding anisotropy in the effective mass tensors.
Rather conflicting results have been reported on the effective masses
as determined by a fitting of IR absorption spectra of the
excited shallow donor states\cite{Suttrop,Gotz} and by
optically detected cyclotron resonance (ODCR).\cite{Son4H,Son6H}

In this letter, calculated
band structures of 6H SiC  and 4H SiC are used to determine
the effective mass tensors at the conduction-band minimum.
Our results are in  good agreement with the ODCR
results on the anisotropies with respect to the c-axis
for 6H and 4H SiC masses and hence
support a mass-anisotropy  as the cause
for the mobility anisotropies. Our calculations  also predict
appreciable anisotropies of the mass tensors in the c-plane.
In addition, they predict
a strong dependency of the masses on band-filling
in the case of 6H SiC due to a double-well-like conduction-band
minimum.

Our  band-structure calculations were performed
within  the density functional theory in the local
density approximation (LDA)\cite{lda}  using
the full-potential (FP) linear muffin-tin orbital
method.\cite{Methfessel,fpdetail}
They differ in some small, but for the present purpose relevant
details, from our earlier results\cite{mrs339} which used
the atomic sphere approximation (ASA).\cite{lmto}
The previous calculations were already
found to provide a satisfactory account of the
UV-reflectivity spectra.\cite{uvsic} In the present work,
we performed calculations on a fine mesh of {\bf k}-points
in order to examine the details
of the energy-band surfaces close to the conduction-band minima.

In  6H SiC, the minimum is found about halfway between $M$ and $L$ along
the $ML\equiv U$-axis of the 6H-Brillouin zone.\cite{bznote}
This leads to a 6-valley model in which the
valleys occur in 3 pairs with each pair being separated by a small
barrier (of 9.5 meV) at the three equivalent $M$ points.
The lowest band  along this line can be well approximated by
a fourth order polynomial in $|k_z|$, with $z$ chosen along ${\bf c}$.
In the directions parallel to  $M\Gamma$
(or $[10\bar{1}0]$) and $MK$ (or $[2\bar{1}\bar{1}0]$), respectively called
the $x$- and $y$-directions,
the bands are found to behave parabolically near the minimum with the
masses there almost equal to those at $M$.
The cyclotron masses for orbits not in the c-plane are calculated from
$m=(\hbar^2/2\pi)\partial A/\partial E$,
with $A$ the cross-sectional area of the extremal orbit on the
constant energy surface calculated numerically from the fitted energy
bands. By varying the energy $E$ of the chosen constant
energy surface, we obtain the variation of the mass with conduction band
filling.

In 4H SiC the minimum is found\cite{Mnote} at $M$  and the bands are parabolic
in the three symmetry directions of  the crystal up to
at least 300 meV above the minimum. In fact, there are two
bands at $M$ within less than 100 meV of each other
whose energetic position was found to be sensitive to
the details of the calculation. Specifically, these bands were interchanged
in the ASA calculation. Apart from this interchange, the results from FP and
ASA calculations were in very good agreement (including the masses
of each of these bands).

The first three rows in Table \ref{t-mass} give
the principal components of the mass tensor at the
$ML$ minimum in 6H SiC and at the $M$ minima in 4H and 2H SiC.
Clearly, these are fully anisotropic (i.e. $m_1\neq m_2\neq m_3$ with
$1,2,3$ corresponding to $x,y,z$.)
This is consistent with the $C_{2v}$ symmetry for points on the $ML$ axis.
2H SiC  is included although its lowest conduction band minimum
is at $K$ (in the equatorial plane at the corner of the hexagon.)
It can be seen that the 2H masses at $M$ are more similar to those of 6H
than those of 4H. Similar closer similarity between 6H and
2H than between 4H and 2H was previously also found in
other aspects of the band structure.\cite{uvsic}
The mass tensor at $K$  is rotationally invariant about the c-axis
as  required by the $C_{3v}$ symmetry of the $K$-point and
has the values: $m\parallel c=0.27$ and $m\perp c=0.45$.

The cyclotron effective mass  associated with a single elliposid is:
\begin{equation}
 m^*=\sqrt{\frac{m_1m_2m_3}{m_1b_1^2+m_2b_2^2+m_3b_3^2}},
\end{equation}
where $b_i$ are the direction cosines of ${\bf B}$
with the principal  axes of the mass tensor.
To compare our results with the ODCR measurements it is necessary to
consider the combined effect of the different valleys
which for a general orientation of the magnetic field ${\bf B}$
become inequivalent.
For ${\bf B}$ along the c-axis, the 3 valleys for 4H
(or double valleys for 6H) at the different $M$-points are equivalently
oriented; consequently
only one peak is observed in ODCR with an effective
mass $m^*_\perp=\sqrt{m_1m_2}$. Our values for this effective mass
are given in the sixth row of Table \ref{t-mass} and are in
good agreement with the ODCR values of $0.42\pm0.02$   in both 4H and 6H SiC.
The slight underestimate of the calculated masses
for 4H is partially due to the
absence of phonon renormalization contributions and may also partially
be due to the LDA. In this connection we note a similar situation in 3C-SiC.
There, our calculated values for the electron mass are
$m_l=0.63$, $m_t=0.23$, while the values measured by cyclotron
resonance\cite{Kaplan} are $m_l=0.677$ and $m_t=0.247$.
(Here  {\sl longitudinal} ($l$) and
{\sl transverse} ($t$) refer to the directions parallel and perpendicular
to the cubic  $\Gamma-X$  axis for the $X$-minimum.)

\begin{figure}
\begin{center}
\mbox{\epsfig{file=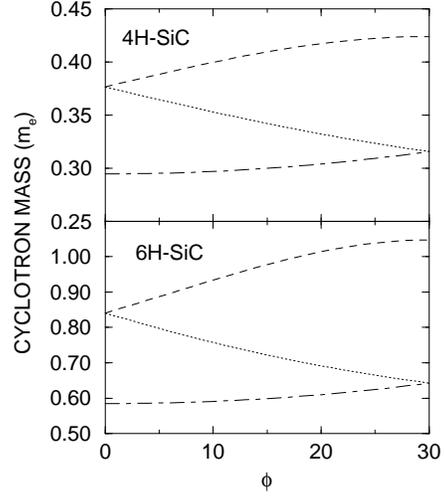,width=10cm}}
\end{center}
\caption{Cyclotron masses in c-plane
for 6H and 4H SiC as function of azimuthal angle $\phi$. $\phi=0$ corresponds
to $[10\bar{1}0]$ and $\phi=30^\circ$  to
$[2\bar{1}\bar{1}0]$. The line type corresponds to the
individual contributions to the ODCR spectra
shown in Fig. \protect{\ref{f-4hq}}.
\label{f-inplan}}
\end{figure}

For ${\bf B}$  in the c-plane, the
ellipsoids at the various M or ML points are generally
oriented differently with respect to the field
and should give rise to three peaks in ODCR.
For the $[10\bar{1}0]$  and  $[2\bar{1}\bar{1}0]$ directions,
two of the peaks coincide. The variation of the masses in the c-plane
is shown in Fig. \ref{f-inplan} as a function of
azimuthal angle $\phi$ measured
from the $[10\bar{1}0]$ direction.

\begin{figure}
\begin{center}
\mbox{\epsfig{file=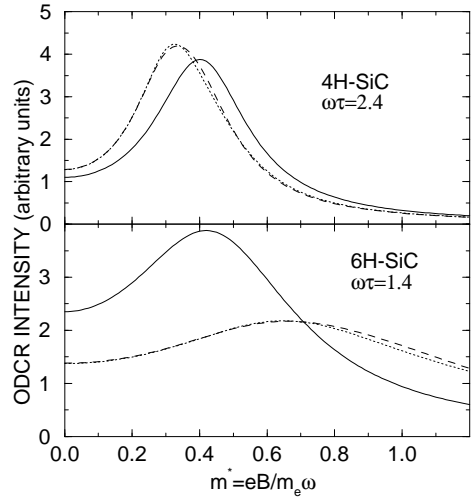,width=10cm}}
\end{center}
\caption{Simulated X-band (9.235 GHz) ODCR spectra for  4H SiC and 6H SiC.
Solid line, ${\bf B} \parallel [0001]$; dashed line,
${\bf B} \parallel [10\bar{1}0]$; dotted line, ${\bf B}\parallel
[2\bar{1}\bar{1}0]$\label{f-mw}}
\end{figure}

The published ODCR measurements  using microwave frequencies
of 9.235 GHz in the X-band do not report anisotropy
in the c-plane and were analysed as if the valleys were
ellipsoids of revolution with a $m_\perp$ and $m_\parallel$,
where $\perp$ and $\parallel$ refer to the c-axis.
Fig. \ref{f-mw}  shows simulated ODCR microwave
absorption spectra  based on the calculated masses for 4H SiC and
6H SiC for a number of important directions. The
$\omega\tau$ values (microwave frequency times lifetime) used are close
to those deduced from the experiments of Son {\sl et al.} \cite{Son4H,Son6H}.
Because of the width  of the ODCR signals
from  the differently oriented ellipsoids, the in-plane sum peak is seen
to show little variation in position. This explains why the
ODCR spectra appear to give an isotropic in-plane  mass in spite
of the fact that each ellipsoid is strongly anisotropic in the plane.
When the effective in-plane masses are equated
to $\sqrt{m^*_\perp m^*_\parallel}$,
as is appropriate for a model with
in-plane rotationally invariant ellipsoids, the
effective $m^*_\parallel$'s in Table \ref{t-mass}  are obtained.

\begin{figure}
\begin{center}
\mbox{\epsfig{file=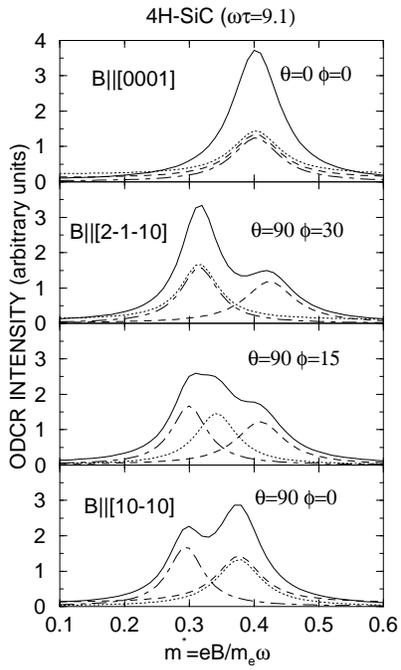,width=10cm}}
\end{center}
\caption{Simulated Q-band (35 GHz) ODCR spectra for 4H SiC
using $\omega\tau=9.1$
for various orientations of the magnetic field. The
dashed, dotted and dot-dashed lines correspond to the individual
effective mass contributions shown in Fig. \protect{\ref{f-inplan}}
and the solid lines to their sums.
For clarity, slight off-sets are included  where
two identical peaks overlap.
\label{f-4hq}}
\end{figure}

Son {\sl et al.} \cite{Son4H} also reported an ODCR signal
using a higher microwave frequency in the Q-band ($\nu=35$ GHz).
With the correspondingly higher
$\omega\tau=9.1$ value, our calculations predict
that the mass anisotropy in the c-plane should be clearly resolved
for 4H, as can be seen in Fig. \ref{f-4hq}.
Very recently, such in-plane anisotropy was confirmed by Hofmann {\sl et al.}
\cite{Hofmann}. The mass values deduced from these new experiments
($m_1=0.58$, $m_2=0.31$ and $m_3=0.33$) are
in good agreement with our calculated values.

For 6H SiC, the strong non-parabolicity resulting from the double well
character of the minima along the $ML$ axis leads to a variation of
the masses with energy above the minimum ($E-E_c$) (or, band filling).
Fig. \ref{f-fill} shows the variation with $E-E_c$ of $\sqrt{m_1m_3}$,
$\sqrt{m_2m_3}$ and the effective $m^*_\parallel$
deduced from the calculated broad peak in microwave  absorption
for ${\bf B}\perp {\bf c}$ in the manner explained above.
The calculated $m^*_\parallel$ is seen to
gradually increase with band filling up to the barrier, at which point
there is a discontinuity because the orbits
then encircle the double minimum centered at $M$ instead of the separate
valleys. At higher energies $m^*_\parallel$ decreases
with increasing energy. In the degenerate limit,
a Fermi energy just at the barrier
corresponds to a carrier concentration of the order of
$10^{19}$ cm$^{-3}$. While carrier concentrations of this magnitude are
certainly possible, the carrier concentrations in the ODCR
measurements\cite{Son6H} (at 6K) which are produced by optical pumping
are expected to be much lower.
Based on the laser intensity employed and the carrier lifetimes
\cite{Son} we estimate concentrations of the order of $10^{15}$cm$^{-3}$.
On the other hand, the high microwave powers used ($>20$ mW) may lead to
a non-equilibrium hot carrier distribution and hence the experiment may
conceivably probe masses higher in the band.

As can be seen in Table \ref{t-mass}, the two calculated 4H masses
and $m_\perp^*$ for 6H SiC are in excellent agreement with experiment.
Also consistent with experiment, $m^*_\parallel$ in 6H SiC
is relatively large. Nevertheless, this mass is 40 \% smaller than experiment.
Several factors, including the non-parabolicity noted above
could bear on this discrepancy. The polaron renormalization mass increase
could be substantial for this large mass  because
the polaron coupling constant increases with mass. Electron-electron
interaction corrections
to the LDA may also alter the masses, but are unlikely to be significant
in view of the good agreement for the other masses. Finally, we note that
the experimental uncertainty on this mass is fairly large because it
depends on fitting a broad resonance.
Considering the above factors, the agreement between the calculated and
measured values for 6H m$^*_\parallel$ may be called fair.

In conclusion, the agreement between our calculated effective mass tensors
and ODCR measurements is excellent for 4H SiC and semiquantitative for
6H SiC. Although we found the effective mass tensors  to be fully anisotropic
in both polytypes, we explain why the recent X-band ODCR measurements
did not resolve the superposition of the fairly broad resonant peaks for
${\bf B}\perp {\bf c}$. That lead to the  erroneous suggestion
that the in-plane mass is isotropic. In agreement with the ODCR studies, we
find the ratio $m^*_\parallel/m^*_\perp<1$ for 4H and $>1$ for 6H SiC.
The in-plane  anisotropy is predicted by our calculations to be observable
with higher $\omega\tau$ values as was confirmed very recently for 4H SiC by
ODCR measurements using a higher microwave frequency.\cite{Hofmann}
We further predict a considerable dependence of
$m^*_\parallel$ in 6H SiC on band filling due to the double well character
of the lowest conduction band. This could influence transport properties
and may also be detectable in future ODCR measurements with
higher band fillings.
The manner in which the effective masses enter the mobility
anisotropy is far from trivial because the different scattering processes
that could come in
have different functional dependence on the masses.
Nevertheless, whatever  scattering  mechanisms are involved,
the mobility is expected to decrease when the mass increases.
The calculated c-axis anisotropies
of the mass tensor thus appear to be the  basis of the
observed large Hall mobility anisotropies in 6H and small and
reversed one in 4H. However,
further detailed work involving  the effects of
different scattering mechanisms will be required to quantitatively
explain their magnitude and temperature dependence.
Finally, it is clear that these rather unusual conduction-band minima
call for an extension of effective mass theory before one can confidently
deduce effective masses from donor excitation IR spectra.

We  thank Drs. W. J. Choyke, R. P. Devaty, N. T. Son,
D. M. Hofmann, D. Volm, and B. K. Meyer for stimulating discussions and
for communicating unpublished results.
This work was supported by the National Science Foundation, Grant No.
DMR-92-22387 and Wright Laboratories through contract No. F33615-93-C-5347.

\newpage
\begin{table}
\caption{Principal components of the conduction-band effective mass tensor
(in units of the free electron mass) for 6H ($ML$-minimum),
4H and 2H SiC (at $M$-minimum) and derived quantities.\label{t-mass}}
\begin{tabular}{ccccc}
      &                       & 6H          & 4H     &  2H \\ \tableline
$m_1$ & $M\Gamma$             & 0.77        &  0.58  & 0.95 \\
$m_2$ & $MK$                  & 0.24        &  0.28  & 0.15 \\
$m_3$ & $ML$                  & 1.42        &  0.31  & 1.07  \\\tableline
$m_{xz}=\sqrt{m_1m_3}$ & $\Gamma-M-L$  & 1.05   & 0.42 &  1.01 \\
$m_{yz}=\sqrt{m_2m_3}$ & $K-M-L$       & 0.59   & 0.29 &  0.40 \\
$m_{xy}=\sqrt{m_1m_2}$ & $\Gamma-M-K$  & 0.43   & 0.40 &  0.38 \\
\tableline
$m^*_\perp=\sqrt{m_1m_2}$ (expt.)\tablenotemark[1]
& ${\bf B}\parallel {\bf c} $ & 0.42     & 0.42 &       \\
$\sqrt{m^*_\perp m^*_\parallel}$ (expt.)\tablenotemark[1] & ${\bf B}\perp{\bf
c}$
& 0.92 & 0.35 & \\
$\sqrt{m^*_\perp m^*_\parallel}$ (theory) &  & 0.67-0.89\tablenotemark[2] &
0.33 &  \\
$m^*_\parallel$ (expt.)\tablenotemark[1]  &   & 2.0$\pm0.2$     & 0.29 &
\\
$m^*_\parallel$ (theory)  &                & 1.1-2.0\tablenotemark[2] & 0.27 &
\\ \tableline
$m^*_\parallel/m^*_\perp$ (expt.)\tablenotemark[1]& & 4.8 &  0.69 & \\
$m^*_\parallel/m^*_\perp$ (theory) & & 2.5-4.6\tablenotemark[2] &  0.68 & \\
\end{tabular}
\tablenotetext[1]{Son et al. \cite{Son4H,Son6H}}
\tablenotetext[2]{Depending on band filling:
first value near bottom of the band, second near barrier crossing}
\end{table}

\newpage

\begin{center}
\mbox{\epsfig{file=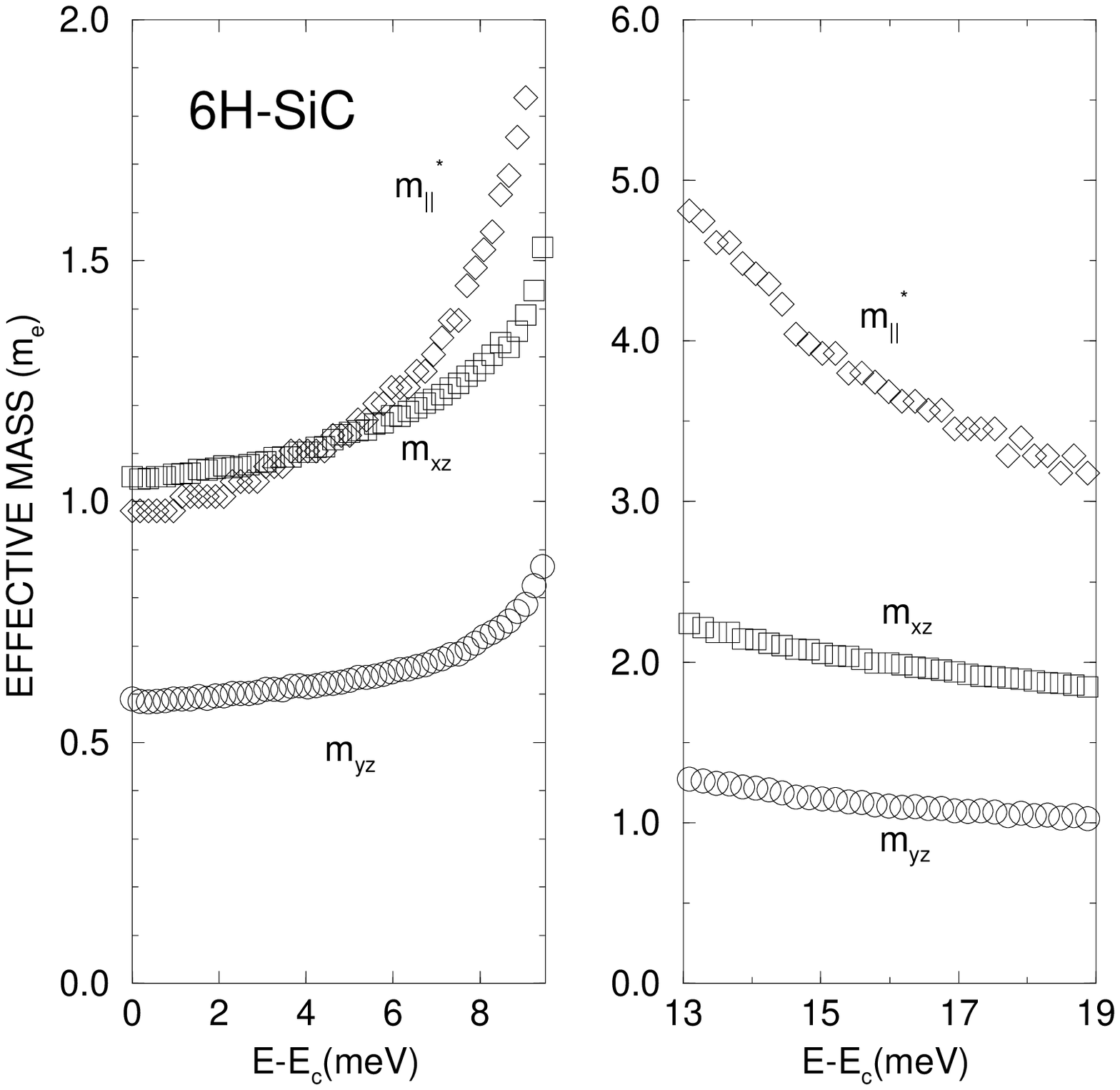,width=10cm}}
\end{center}
\begin{figure}
\caption{Calculated effective masses in 6H SiC as a function of
conduction band filling. The six-valley  region below the barrier
is shown separately from the three-valley region above.
$m_{xz}$ and $m_{yz}$ are defined in
Table I. $m^*_\parallel$ is a pseudo-effective mass parallel to $c$
appearing in the isotropic in-plane  model
discussed in the text.\label{f-fill}}
\end{figure}

\begin{references}
\bibitem{Schaffer} W. J. Schaffer, G. H. Negley, K. G. Irvine, and
J. W. Palmour, in  {\sl Diamond, SiC and Nitride Wide Bandgap Semiconductors},
edited by C. H. Carter, Jr., G. Gildenblat, S. Nakamura, and R. J. Nemanich,
Mater. Res. Soc. Symp. Proc. Vol. 339, p. 595 (1994).
\bibitem{Schadt} M. Schadt, G. Pensl, R. P. Devaty, W. J. Choyke, R. Stein,
and D. Stephani, {\sl Appl. Phys. Lett.} {\bf 65}, 3120 (1994).
\bibitem{Suttrop} W. Suttrop, G.  Pensl, W. J. Choyke,
R. Stein, and S. Leibenzeder, {\sl J. Appl. Phys.} {\bf 73}, 3708 (1992)
\bibitem{Gotz} W. G\"otz, A. Sch\"oner, G. Pensl, W. Suttrop, W. J. Choyke,
R. Stein, and S. Leibenzeder, {\sl J. Appl. Phys.} {\bf 73}, 3332 (1993)
\bibitem{Son4H} N. T. Son, W. M. Chen, O. Kordina, A. O. Konstantinov,
B. Monemar, E. Janz\'en, D. M. Hofmann, D. Volm, M. Drechsler, and B. K. Meyer,
{\sl Appl. Phys. Lett.} {\bf 66}, 1074 (1995).
\bibitem{Son6H}  N. T. Son,  O. Kordina, A. O. Konstantinov, W. M. Chen,
E. S\"orman, B. Monemar, and E. Janz\'en, {\sl Appl. Phys. Lett.} {\bf 65},
3209 (1994).
\bibitem{lda} P. Hohenberg and W. Kohn, {\sl Phys. Rev.} {\bf 136},
B864 (1964); W. Kohn and L. J. Sham, {\sl Phys. Rev.} {\bf 140}, A1133 (1965);
L. Hedin and B. I. Lundqvist, J. Phys. C {\bf 4}, 2064 (1971).
\bibitem{Methfessel} M. Methfessel, {\sl Phys. Rev. B} {\bf 38}, 1537 (1988).
\bibitem{fpdetail} The calculations were performed at the experimental
$a$ lattice constant using an ideal $c/a$ ratio. A triple $\kappa$
$dds$ basis set was used on Si and C nearly touching muffin-tin spheres
and additional s-orbitals were used on some of the empty spheres
in the case of 6H SiC.
Augmentation and fitting cut-offs were $l_{max}=4$ on atoms
and $l_{max}=2$ on empty spheres.  12 special k-points
were used in the Brillouin-zone integrations.
Complete details and results of these calculations,
which were found to be adequately converged with respect to all
relevant parameters, will be presented elsewhere.
\bibitem{mrs339} W. R. L. Lambrecht
in {\sl Diamond, SiC and Nitride Wide Bandgap Semiconductors},
edited by C. H. Carter, Jr., G. Gildenblat, S. Nakamura, and R. J. Nemanich,
Mater. Res. Soc. Symp. Proc. Vol. 339, p. 565 (1994).
\bibitem{lmto} O. K. Andersen, O.~Jepsen, and M.~\v Sob,
in {\it Electronic Band Structure and its Applications}, edited
by M.~Yussouff, (Springer, Heidelberg, 1987) p. 1.
\bibitem{uvsic} W. R. L. Lambrecht, B. Segall, M. Yoganathan,
W. Suttrop, R. P. Devaty, W. J. Choyke, J. A. Edmond, J. A. Powell,
and M. Alouani, {\sl Phys. Rev. B} {\bf 50}, 10722  (1994).
\bibitem{bznote} The $M$  point lies in the equatorial plane ($k_z=0$) at
the midpoint of the hexagon sides, while $L$ lies above it in the top plane
($k_z=\pi/c$); see e.g. C. J. Bradley and A. P. Cracknell,
{\sl The Mathematical Theory of Symmetry in Solids:
Representation Theory for Point Groups and Space Groups},
(Clarendon Press, Oxford 1972).
\bibitem{Mnote} The location of the minimum at $M$ in 4H SiC  has been
under debate because of the claim by
L. Patrick, W. J. Choyke, and D. R. Hamilton,
{\sl Phys. Rev.} {\bf 137}, A1515 (1965) that phonon-replica in luminescence
spectra indicate that the minima occur at lower symmetry  $F$-points.
We will show elsewhere  that the spectrum is in fact consistent with
a $M$-minimum. This demonstration utilizes the recent phonon calculations
by  M. Hofmann, A. Zywietz, K. Karch, and F. Bechstedt,
{\sl Phys. Rev. B} {\bf 50}, 13401 (1994).
In addition, the present results for the masses indirectly confirm
the location of the minimum at $M$.
\bibitem{Patrick} L. Patrick, {\sl Phys. Rev. B} {\bf 5}, 2198 (1972).
\bibitem{Kaplan} R. Kaplan, R. J. Wagner, H. J. Kim, and R. J. Davis,
{\sl Solid State Commun.} {\bf 55}, 67 (1985).
\bibitem{Hofmann} D. M. Hofmann, D. Volm, and B. K. Meyer,
private communication.
\bibitem{Son} N. T. Son, private communication.
\end{references}
\end{document}